\documentclass[lineno]{jfm}

\usepackage{psfrag,latexsym,amssymb,amsmath,natbib,color,epstopdf}
\usepackage{graphicx}[dvips]

\def\Real{{\, \rm{Re}}}
\def\Imag{{\, \rm{Im}}}
\def\dd{{\, \rm{d}}}
\def\dr{{\rm{d}}}

\def\bra{\langle}
\def\ket{\rangle}
\def\p{\partial}

\def\beq{\begin{equation}}
\def\eeq{\end{equation}}
\def\la{\label}
\def\ii{{\rm i}}

\def\r#1{(\ref{#1})}
\newcommand{\mylab}[3]{\raisebox{#2}[0mm][0mm]{%
\makebox[0mm][l]{\hspace*{#1}{#3}}}}%
\def\utau{u_\tau}
\def\retau{Re_\tau}

\def\hu{\widehat{u}}
\def\hp{\widehat{p}}
\def\hv{\widehat{v}}

\def\homega{\widehat{\omega}}
 
\newcommand\Rey{\mbox{\textit{Re}}}  

\def\spacce#1{\hskip #1pt}
\def\drawline#1#2{\raise 2.5pt\vbox{\hrule width #1pt height #2pt}}
\def\solid{\drawline{24}{.5}\nobreak}

\def\bdash{\hbox{\drawline{5.8}{.5}\spacce{2}}}

\def\dashed{\bdash\bdash\bdash\nobreak}

\def\chndot{\hbox%
{\drawline{4.6}{.5}\spacce{2}\drawline{1}{.5}\spacce{2}\drawline{4.6}{.5}\spacce{2}\drawline{1}{.5}\spacce{2}\drawline{4.6}{.5}}\nobreak }

\def\trian{\raise 1.25pt\hbox{$\scriptstyle\triangle$}\nobreak}

\def\dtrian{\raise 1.25pt\hbox%
{$\scriptscriptstyle\bigtriangledown$}\nobreak}

\def\squar{\raise 1.25pt\hbox{$\scriptstyle\Box$}\nobreak}

\def\diamon{\raise 1.25pt\hbox{$\scriptstyle\diamond$}\nobreak}

\newcommand{\soliddtrian}{$\blacktriangledown$\nobreak}

\def\linedtri1{\hbox{\bdash\hspace{-1.6mm}$\bigtriangleup$\hspace{-0.8mm}\bdash}\nobreak}
\def\soliddtrian1{$\blacktriangledown$\nobreak}
\def\solidrtrian2{$\blacktriangleright$\nobreak}
\def\solidltrian3{$\blacktriangleleft$\nobreak}

\title{Wall-bounded turbulence needs not be long}

\author{Carlos Mart\'inez--L\'opez and Javier Jim\'enez}
\affiliation{School of Aeronautics, U. Polit\'ecnica Madrid, 28040 Madrid Spain}
\date{\today}

\begin{document}
\maketitle
\begin{abstract}
Experiments on the regeneration of long streaks in flows in which they had originally been damped show that their initial growth is due to the interaction of the mean shear with weak long cross-flow velocities (`rollers') that remain even when the streaks are damped. More surprisingly, turbulence also persists in simulations in which only the long rollers are damped while long streaks remain, and these flows are also able to recover when the damping is removed. Finally, simulations are presented in which both streaks and rollers longer than $\lambda_x^+\approx 600$ are damped. They survive and regenerate, and an interpretation in terms of their energy balance is provided. In contraposition to the classical minimal channels, which include infinitely long structures, these new flows do not contain any feature longer than the damping wavelength, and support a model in which wall turbulence only depends on processes for which the geometric $\lambda_x-\lambda_z$ aspect ratio is of order unity.
\end{abstract}


\section{Introduction}

Very large numbers in physics typically imply the coexistence of separate processes connected by a matching principle that becomes asymptotically exact in the limit of infinite separation \citep[see e.g.][]{cole68}. Turbulence is no stranger to this segregation: energy is generated at large scales and dissipates at the smallest ones, and one of the central objects in turbulence research is the cascade connecting the two regimes \citep{kol41}. Wall-bounded turbulence provides an additional example of momentum transfer across a hierarchy of attached eddies \citep{tow:61}. In both cases, the ratio between the inner and outer scales is proportional to some power of the Reynolds number that provides a potentially infinite expansion parameter.

Less often discussed is the elongated geometry of the streamwise velocity streaks \citep{kli:rey:sch:run:67}. They are generally believed to originate from the deformation of the mean velocity profile by the wall-normal velocity \citep{kim:kli:rey:71}, but the two-point correlation of the latter is much shorter than that of the former \citep{jim18}, and the reason for the discrepancy is not well understood. Admittedly, this is a less natural asymptotic problem than the two examples above, because the ratio of the two correlation lengths in boundary layers and channels is at most of the order of 10--20 \citep{ sil:jim:mos:14}, but Couette flow has much longer streaks \citep{PirBerOrl14,lee:moser:18}, and the existence of flows with similar wall-normal velocity structures \citep{sil:jim:mos:14} and different streamwise-velocity streaks suggest that the two processes have different origins, and that it may be useful to study the connecting mechanism.

The small-scale limit of this relation has been studied the most; meandering instabilities with wavelengths of the order of the width of the streaks create bursts with aspect ratios of order unity \citep{kim:kli:rey:71,jim:moi:91,Schoppa02}, but the route from short bursts to long streaks has been explored less.
 
A useful model was introduced by \cite{jim:moi:91}, who showed that a minimal self-sustaining unit of the buffer layer of plane Poiseuille turbulence contains one or two short bursting vortices and a streak that is infinitely long when compared to the computational box. \cite{jim:pin:99} modified different terms of the evolution equations of such a unit. They showed that both streaks and vortices are required, and numerous minimal kinematic and dynamical models have resulted in a widely accepted minimal regeneration cycle \citep{jim94,Hamilton95,Waleffe97}. This was extended to the logarithmic layer by \cite{flo:jim:10}, but the length of these minimal units is never much greater than their spanwise width.

An obvious possibility is that streaks are wakes left in the mean shear by differentially advected bursts \citep{ala:jim:zan:mos:04}. Optimal initial conditions for the transient linearised growth of perturbations of the mean velocity profile of a turbulent channel result in infinitely long streaks and shorter bursts \citep{but:far:93,ala:jim:06,jim:13a}, although this is mostly because longer features decay more slowly. \cite{vaug:etal:15}, among others, have presented plausible self-sustaining models of wall turbulence based on time-dependent infinitely long streaks, but this optimal amplification also requires an infinitely long time \citep{ala:jim:06}, and \cite{del:jim:zan:mos:06} noted that the lifetime of individual bursts is too short for their wakes to explain the observed streak length. The problem of creating long streaks becomes that of creating long packets of bursts.

In fact, there is ample evidence that long streaks are associated with several bursts. \Citet{del:jim:zan:mos:06} and \cite{loz:flo:jim:12} showed that the down- and up-draft components of bursts are statistically found in pairs associated with quasi-streamwise vortices located along the edges of streamwise-velocity streaks that organise them longitudinally. \cite{adr:mei:tom:00} and \cite{adr07} propose a plausible model in terms of `hairpin' packets, and demonstrate the formation of short packets in the absence of background fluctuations. Composite packets in pipes are described by \cite{bal:adr:Wu:13}, organised by large-scale weak streamwise rolls. However, most of these works discuss correlation, rather than causation, and it is unclear whether the burst packets create the streaks or vice versa. For example, although wall-bounded flows clearly contain long streaks, \cite{jim22_nostr} showed that they are not essential features of wall turbulence, which only requires that fluctuations of the streamwise velocity have lengths comparable to the bursts themselves.
 
A possibility is that the causal scale interactions take place across different wall distances, because a burst in the outer part of the boundary layer is large when compared to the width of the near-wall streaks, and the footprints of these outer structures reach the wall \citep{tow:61,hoy:jim:06,hut:mar:07}. Conversely, \cite{TohItan05} proposed that the outer streaks are created by a `bunching' instability of near-wall ones.

The present paper deals with the first stages of the process by which streaks in real turbulence get to be much longer than the bursts that presumably create them. The strategy will be to start with initial conditions in which long streaks have been inhibited and observe how they evolve after the inhibition is removed. Besides providing an observational setup relatively free of other perturbations, this technique incorporates a time constraint. It is often noted that weak rolls can create strong streaks, but they need a long time to do it, and the regeneration allows us to estimate that time.

In addition, explaining streak regeneration motivates us to study the effect of shortening other flow features, with the surprising conclusion that wall-bounded turbulence requires essentially no input from the long scales. There exists a functional turbulence `engine' with a wall-parallel aspect ratio of order unity, and the resulting flow shares many of the features of full turbulence.

The organisation of the paper is as follows. The numerical setup is described in \S\ref{sec:simul}. Results of the streak regeneration experiments are presented in \S\ref{sec:results}, and the new short-flow simulations are discussed in \S\ref{sec:short}. Conclusions are offered in \S\ref{sec:conc}.
\begin{table}
  \begin{center}
    \def~{\hphantom{0}}
    \begin{tabular}{llccccccccc}
      Case  &  goes to &$U_b h/\nu$ &  $Re_\tau$ & $(\Lambda_x \times  \Lambda_z) $ & $\Delta x^+$ &$\Delta z^+$ & 
          $\Delta y^+_{max}$ & $\lambda_{f}/h$ & $\lambda_{f}^+$ & $(x,y,z)$ grid points\\[3pt]
      F550 & SS550:10/15        & $10060$ & 550& $(8\times 4)\pi h$ & 8.9 & 5.5 & 6.7  & $\infty$  & $\infty$  & $1536\times 257\times 1536$\\
      SS550:10 & RS550:10       & $10060$ & 452 & $(8\times 4)\pi h$ & 7.4 & 3.7 & 5.5  & 2.28 & 1033 & $1536\times 257\times 1536$\\
  SS550:15 & RS550:15       & $10060$ & 381 & $(8\times 4)\pi h$ & 6.2 & 3.1 & 4.7  & 1.57 & 598   & $1536\times 257\times 1536$\\
      F180 & S*180        & $2920$ & 186 & $(12\times 4)\pi h$ & 9.1 & 4.6 & 6.1  & $\infty$  & $\infty$  &  $768\times 97\times 512$\\
      SS180   & RS180     & $2920$ & 150 & $(12\times 4)\pi h$ & 7.4 & 3.7 & 4.9  & 3.77  & 569  & $768\times 97\times 512$\\
      SA180   & RA180     & $2920$ & 170 & $(12\times 4)\pi h$ & 8.4 & 4.2 & 5.6  & 3.77  & 643  & $768\times 97\times 512$\\
      SR180:5 & RR180:5   & $2920$ & 179 & $(12\times 4)\pi h$ & 8.8 & 4.4 & 5.9  & 7.54  & 1351 & $768\times 97\times 512$\\
  \end{tabular}
\caption{Parameters of the reference and damped simulations used as initial conditions for the recovery experiments. 
The computational box is $\Lambda_x\times \Lambda_z$, and the grid dimensions and resolutions, $\Delta *$, are given in terms of collocation points, using the friction velocity of the short-streak simulation whose shortest damped wavelength is $\lambda_{f}$. The asterisk in the fourth row of the table stands for S (streak), R (roller) or A (all), depending on the long features being damped (see \S \ref{sec:short}). At the end of a full recovery, the value of $Re_\tau$ and the rest of statistics return to be those of F550 or F180, depending on the case.
 }%
\label{tab:cases}
\end{center}
\end{table}

\section{Numerical experiments}\la{sec:simul}

We analyse simulations of pressure-driven incompressible turbulent channel flow between smooth parallel plates separated by $2h$, in computational boxes periodic in the streamwise $(x)$ and spanwise $(z)$ directions. The code integrates the evolution of $\nabla^2 v$ and the wall-normal vorticity, $\omega_y$, using dealiased Fourier in $(x,z)$ and Tchebychev in the wall-normal, $y$, direction \citep{KMM87}. Other variables are obtained from continuity, and the volume flux per unit span, $2hU_b$, is kept constant. The velocity components along $(x,y,z)$ are $(u,v,w)$, respectively. Capitals refer to quantities averaged over wall-parallel planes, denoted by $\bra\,\ket$, and lower-case symbols are fluctuations with respect to those averages. Primes are root-mean-squared fluctuation intensities.
 
\cite{jim22_nostr} used this code to run simulations in which the spanwise variation of the streamwise velocity is damped for wavelengths longer than a chosen limit $\lambda_{f}$, including the infinitely long $\lambda_x=2\pi/k_x=\infty$. Following \cite{jim:pin:99}, this is implemented by zeroing at each time step all the long harmonics of the Fourier expansion of $\omega_y$, which is approximately equal to $\p_z u$ for long and narrow features. Specifically, if $\omega_y (x,y,z) =\sum\sum\homega_y (k_x,y,k_z) \exp[\ii (k_x x +k_z z)]$, we force $\homega_y (k_x,y,k_z) = 0$ for all $|k_z|>0$, and $|k_x|\le 2\pi/\lambda_{f}.$ The $\nabla^2\hv$ equation is not modified in these experiments, but we will later discuss simulations in which the long wavelengths of $\nabla^2\hv$ (denoted by $\nabla^2\hv_L$) are damped, by themselves or together with $\omega_y$. This is equivalent to zeroing $\hv$. Numerical parameters are summarised in table \ref{tab:cases}, and more details are found in \cite{jim22_nostr}, and in \cite{juanc03}, whose undamped simulations are used as reference (F550 and F180 in table \ref{tab:cases}). A new truncated simulation at lower Reynolds number (SS180 in table \ref{tab:cases}) was run to explore the scaling of the interactions between the near-wall and outer flow, and as initial condition for the $\nabla^2 v$-damped recoveries. Because one of the effects of shortening the streaks is to change the friction velocity, $\utau$, we use two normalisations: wall units defined with the `instantaneous' friction velocity of the evolving flow, $u_\tau$, are denoted by a `$+$' superscript, while those using the friction velocity of the F180 or F550 undamped simulations, distinguished by a `0' subscript, are denoted by `$\times$'. The eddy turnover used to normalise time is always $h/u_{\tau 0}$.

The recovery experiments use the same code without damping. Each simulation is initialised from a converged field in which $\omega_y$ or $\nabla^2 v$ has been damped, and the flow is observed as the damped wavelengths are re-established. Notationally, a recovery simulation initialised from case SS** (short streak) is denoted as RS** (recovering streak), with additional notation explained in the caption of table \ref{tab:cases}.

In addition, because elongated features are mostly absent from the central part of the channel, where the velocity fluctuations are fed by turbulent transport from layers closer to the wall \citep{hoyas08}, the following analysis is restricted to $y/h\le 0.8$. We will often use `streak' and `roller' as shorthand for streamwise-long fluctuations of the streamwise or cross-flow velocity, $u^2_L$ and $R^2_L=v^2_L+w^2_L$, respectively. Averages over $y$ are denoted by $\overline{(*)} = \int_0^h *(y) \dd y/h$, and $q^2=u^2+R^2$.

\section{Recovery of streaks}\la{sec:results} 

\begin{figure}
\centerline{%
\raisebox{0mm}{\includegraphics[height=.20\textwidth,clip]%
{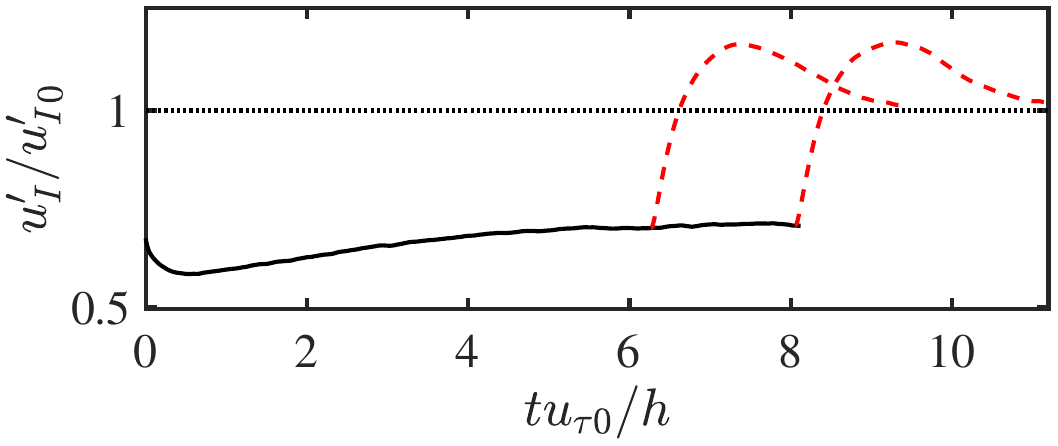}}
\mylab{-.49\textwidth}{.18\textwidth}{(a)}%
\hspace*{4mm}%
\raisebox{1mm}{\includegraphics[height=.20\textwidth,clip]%
{ 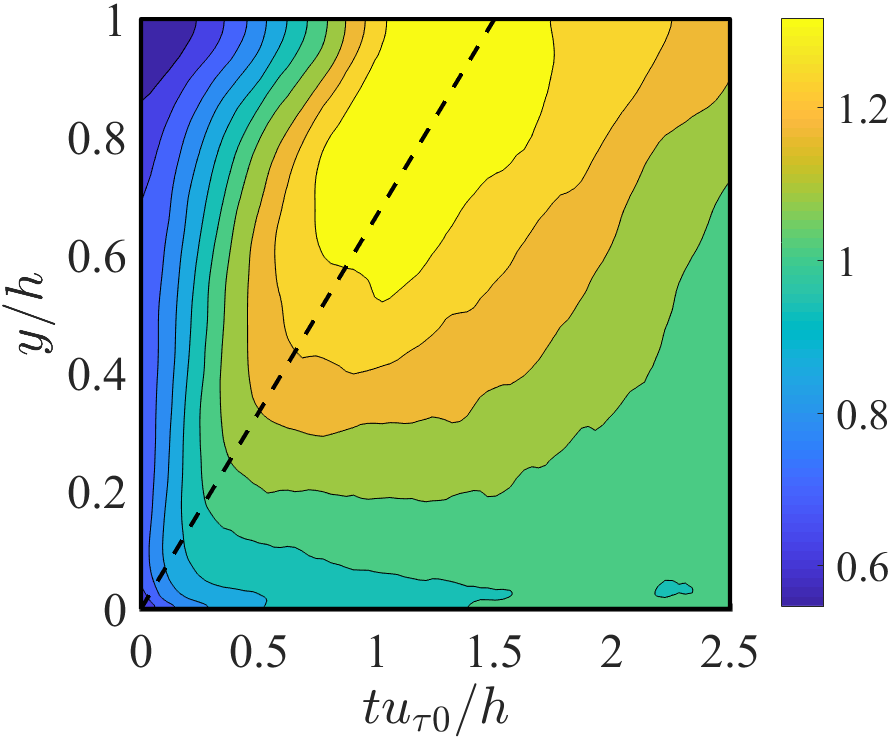}}
\mylab{-.265\textwidth}{.18\textwidth}{(b)}%
\hspace*{5mm}%
\raisebox{1mm}{\includegraphics[height=.20\textwidth,clip]%
{ 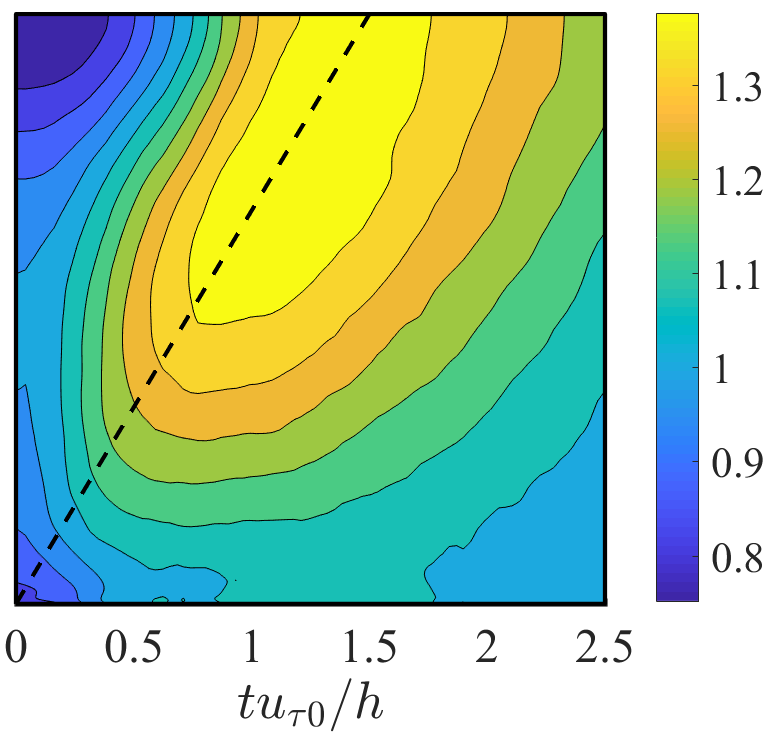}}
\mylab{-.25\textwidth}{.18\textwidth}{(c)}%
}%
\caption{%
(a) Evolution of the integrated intensity of the streamwise velocity fluctuations,
$u'_I=(\overline{u'^2})^{1/2}$, in the damping (SS550:10, solid line) and recovery experiments
(RS550:10, dashed), normalised with its equilibrium value in a canonical channel.
(b) Streamwise-velocity fluctuation profile versus time, $u'/u'_0$. 
(c) Transverse-velocity, $R'/R'_0$. Case RS550:10 normalised with the profile of the reference undamped simulation, F550. The dashed diagonal is $\dr y/\dr t=1.5 u_{\tau 0}$.
}
\label{fig:cfs}
\end{figure}

Figure \ref{fig:cfs}(a) displays the history of a typical streak-recovery experiment (RS550:10). Starting from an undamped reference channel (F550), the streaks are damped below the 10th wavenumber, causing an instantaneous drop of the streamwise velocity fluctuations. The flow is allowed to settle to its new short-streak equilibrium (SS550:10) before the damping is released and the flow returns to its reference state. For each of the damped simulations, we run two recovery experiments starting from damped flow fields separated by $t u_{\tau 0}/h \approx 2$. They behave similarly, and the data discussed below are averaged over these two simulations. Except for this figure, $t=0$ refers to the moment at which the damping is released.

Figure \ref{fig:cfs}(b) shows the evolution of the velocity fluctuation profile during the recovery. It initially overshoots, as in figure \ref{fig:cfs}(a), moves away from the wall and eventually relaxes towards equilibrium. This outwards drift could be interpreted as that the new fluctuations are advected by the wall-normal velocity, but the implied advection velocities, $1.5u_{\tau 0}$ (approximately $1.8 u_\tau$ at $t=0$), are higher than previously observed values, which are closer to $u_\tau$ \citep{flo:jim:10,loz:jim:14}. An alternative explanation is that the time scale of the recovery is determined at each $y$ by the local shear, which also leads to an apparently constant advection velocity,
\beq
t_{rec}\sim S^{-1}(y)\approx \kappa y/u_\tau\quad \Rightarrow \quad 
\dr y/\dr t\approx y/t_{rec} \approx u_\tau/\kappa \approx 2.5 u_\tau,
\la{eq:St}
\eeq
where $\kappa \approx 0.4$ is the von K\'arm\'an constant. 

\begin{figure}
\vspace*{1mm}%
\centerline{%
\raisebox{0mm}{\includegraphics[width=.30\textwidth,clip]%
{ 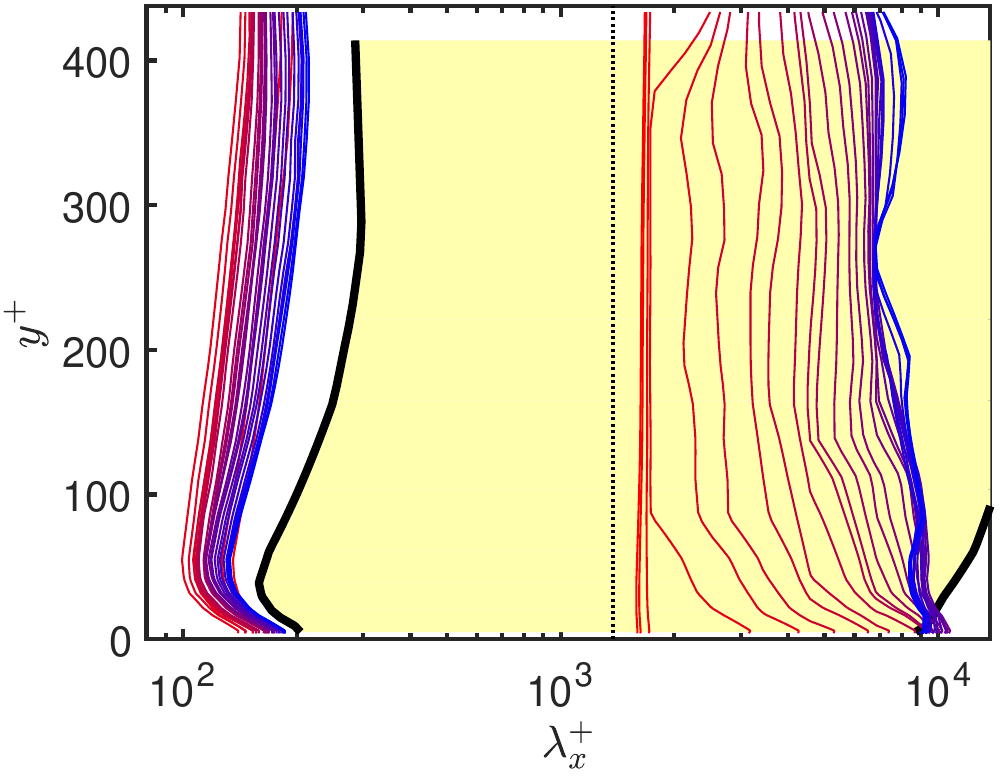}}
\mylab{-.325\textwidth}{.215\textwidth}{(a)}%
\hspace*{3mm}%
\raisebox{0mm}{\includegraphics[width=.30\textwidth,clip]%
{ 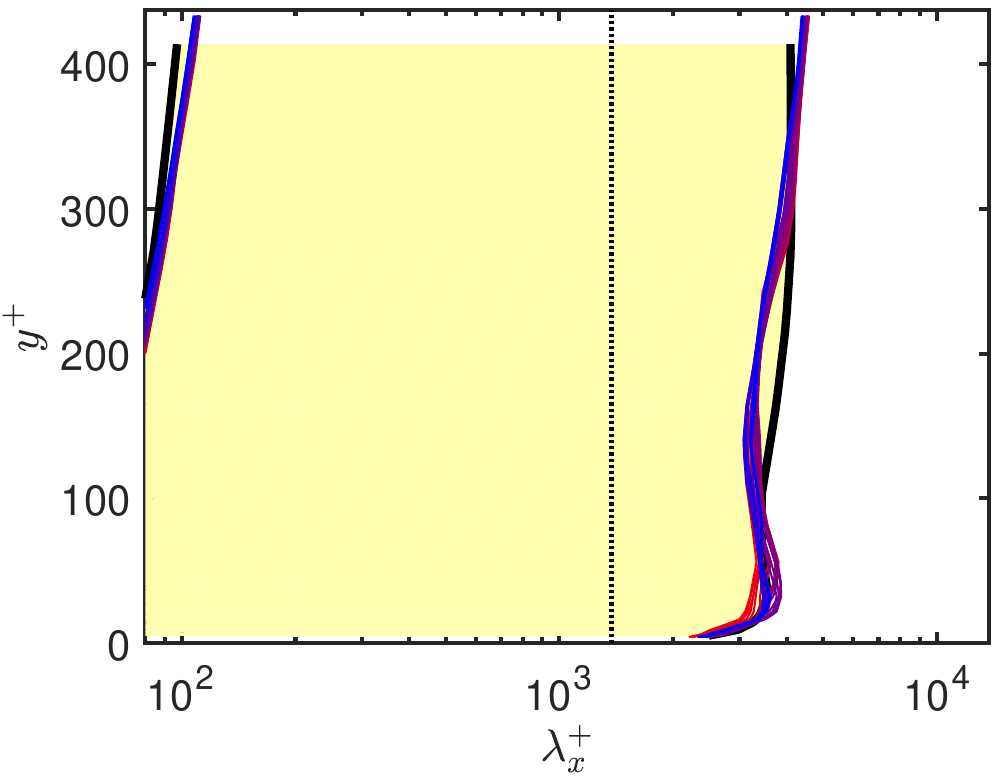}}
\mylab{-.325\textwidth}{.215\textwidth}{(b)}%
\hspace*{3mm}%
\raisebox{0mm}{\includegraphics[width=.30\textwidth,clip]%
{ 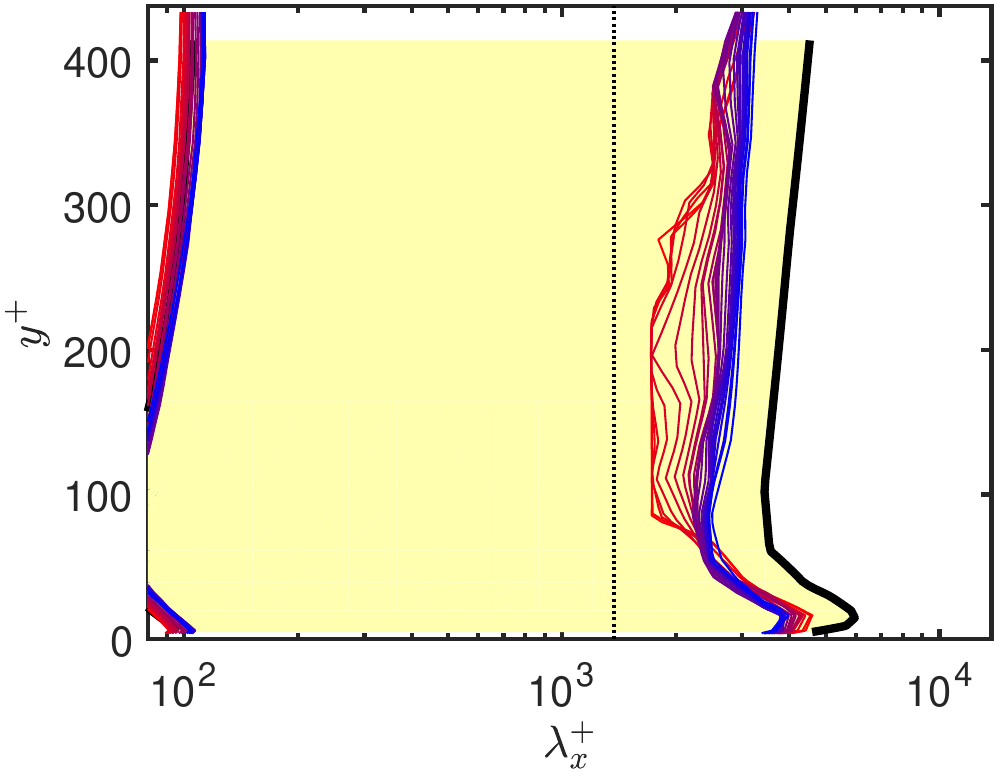}}
\mylab{-.325\textwidth}{.215\textwidth}{(c)}%
}%
\caption{%
(a) Streamwise one-dimensional premultiplied spectrum of the streamwise velocity, $k_x E_{uu}(k_x)/u'^2$, as a function of time and $y$ in RS550:10. Contours are $t u_{\tau 0}/h=0 (0.0044) 0.337$, from red to blue, and contain 90\% of the energy of the spectrum.
(b) As in (a) for $k_x E_{vv}/v'^2$.
(c) $k_x E_{ww}/w'^2$.
The thick black contour with yellow fill is the spectrum of the reference flow, F550, and the vertical dotted line is the damping wavelength.
}
\label{fig:spec}
\end{figure}

The spectra in figure \ref{fig:spec} show more details of the elongation process. Figure \ref{fig:spec}(a) is the evolution of the streamwise velocity spectrum during the initial stages of the regeneration. The streaks grow fastest near the wall, where they have almost reached their equilibrium length slightly before contours become purple after 0.1 turnovers. The outer velocity lengthens more slowly, and does not immediately influence the wall. This is confirmed by the regeneration experiments RS180 at a lower Reynolds number (not shown), which essentially has no outer flow. Its near-wall region behaves similarly to the RS550s, and there are few differences between the two Reynolds numbers. Figures \ref{fig:spec}(b,c) show that the growth of the near-wall $u$ is not accompanied by that of the transverse velocity components, whose length has changed little even after the streamwise velocity has grown to half its final length.

\begin{figure}
\centerline{%
\raisebox{0mm}{\includegraphics[height=.245\textwidth,clip]%
{ 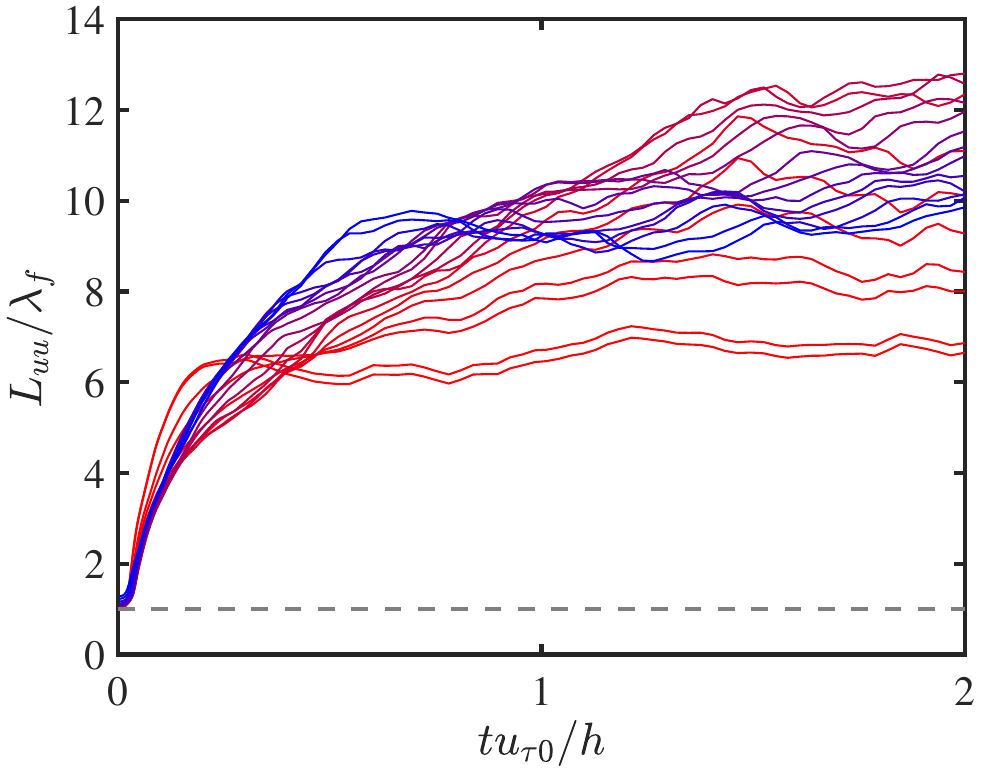}}
\mylab{-.24\textwidth}{.21\textwidth}{(a)}%
\hspace{1mm}%
\raisebox{0mm}{\includegraphics[height=.245\textwidth,clip]%
{ 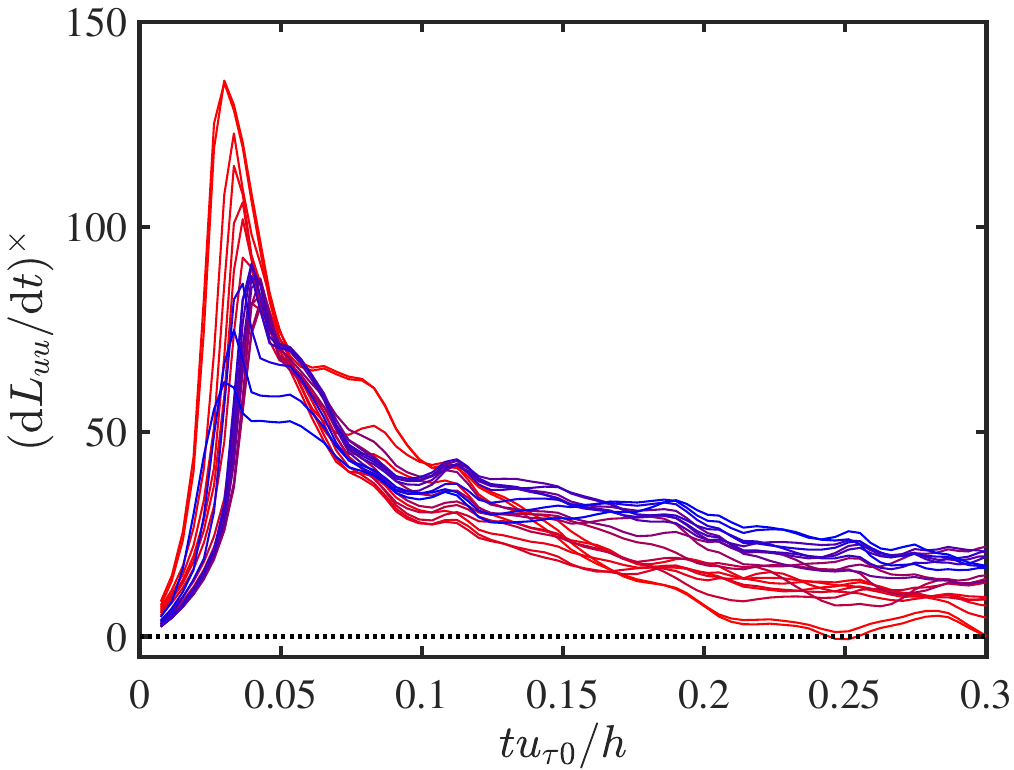}}
\mylab{-.06\textwidth}{.21\textwidth}{(b)}%
\hspace*{1mm}%
\raisebox{0mm}{\includegraphics[height=.245\textwidth,clip]%
{ 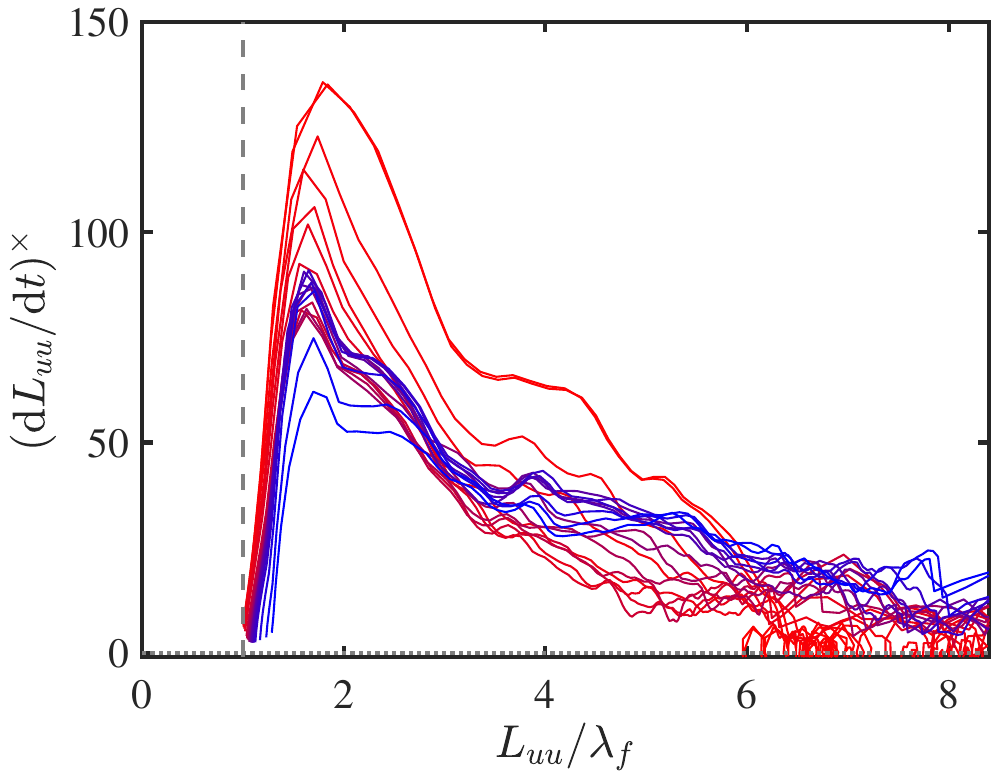}}
\mylab{-.06\textwidth}{.21\textwidth}{(c)}%
}%
\caption{%
(a) Evolution of the wavelength containing 85\% of the energy in the streamwise spectra during recovery. RS550:15. From red to blue, $y^\times=4$ to $y/h=0.8$. The dashed horizontal line is the initial $\lambda_f$.
(b) Elongation velocity $\dr L/\dr t$ versus recovery time and wall distance. 
(c) As in (b), versus streak length. 
}
\label{fig:Lx}
\end{figure}

\begin{figure}
\raisebox{0mm}{\includegraphics[width=\textwidth,clip]%
{ 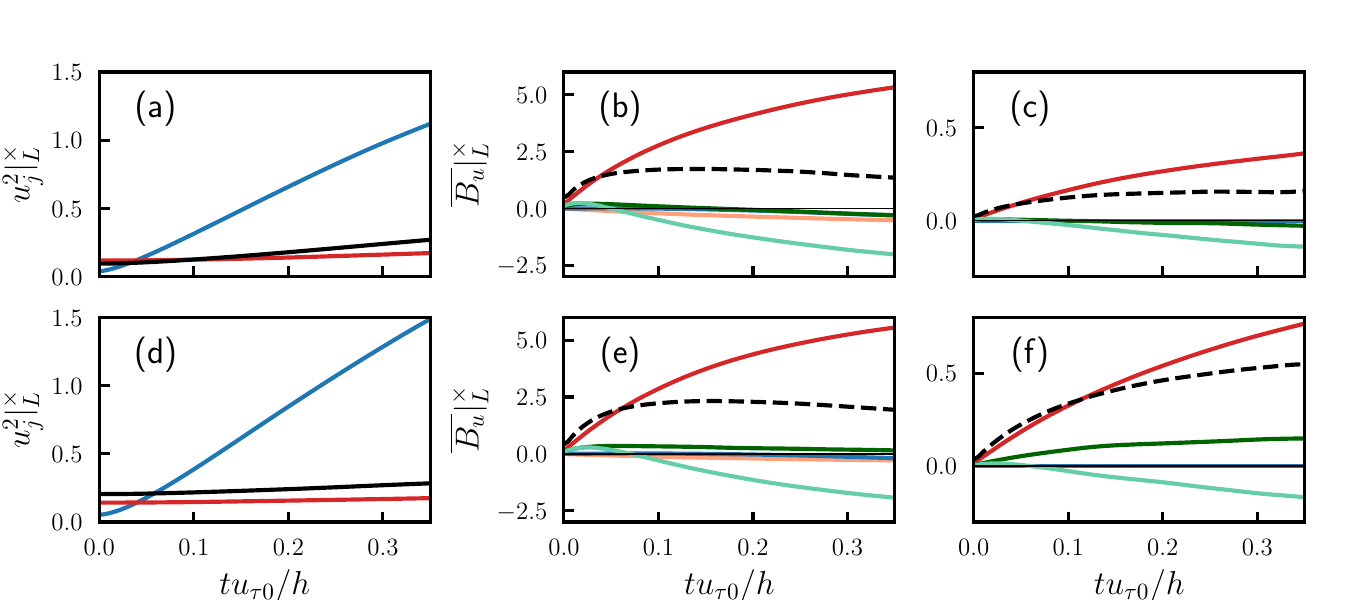}}
\caption{%
(a) Evolution of the integrated energy of the long $(\lambda_x\ge\lambda_f)$ velocity components during the recovery RS550:15. Blue, $\overline{u^{2}}|_L^\times$; red, $\overline{v^{2}}|_L^\times$; black, $\overline{w^{2}}|_L^\times$.
(b) Dominant source terms of the energy equation for $\overline{u^{2}}|_L$ versus recovery time.
Red, production from the mean profile, $-\Real\overline{\bra \hu \hv^*\ket_L \p_yU}$; 
green, production from the streak fluctuations, $- \Real\overline{\bra\widehat {u v}\,\p_y\hu^*\ket}_L$; 
cyan, spanwise streak advection, $\Imag\overline{\bra\widehat {u w}\, k_z \hu^*} \ket_L$; 
blue, streamwise streak advection, $\Imag\overline{\bra\widehat {u u}\, k_x \hu^*} \ket_L$;  
orange, pressure-strain, $-\Imag\overline{\bra \hp^*k_x \hu} \ket_L$;  
black,  $\p_t \overline{u^{2}}|_L/2$ \citep{mizuno16}. All terms are summed over long wavenumbers and averaged over $y$ across the channel half-height $h$. Dissipation is omitted.
(c) As in (b) for $E_{uu}(k_x=0, |k_z|>0)$.
(d--f) As in (a--c) for RS180.
}
\label{fig:balu}
\end{figure}

Figure \ref{fig:Lx}(a) displays the evolution of the spectral length of the streamwise velocity, $L_{uu}$, defined as the wavelength containing a given fraction of the energy ($\alpha=0.85$ in the figure, but the conclusions hold for $\alpha\in [0.6-0.9]$),
\beq
\int_{2\pi/L_{uu}}^\infty E_{uu}(k_x) \dd k_x = \alpha u'^2,
\la{eq:Lx}
\eeq
with natural extensions to other velocity components and to the cospectrum. This definition approximately agrees with the right-hand limit of the spectra in figure \ref{fig:spec} and with the truncation length. Note that $L_{uu}(t=0)\approx \lambda_f$ in figure \ref{fig:Lx}(a). It is thus more convenient for our purpose that the more usual correlation length $\ell=\int_0^\infty R(x) \dd x= (2/\pi u'^2)\,E_{uu}(k_x=0)$, where the two-point correlation $R$ is the Fourier transform of $E_{uu}$. This length measures the energy in the $k_x=0$ mode and vanishes identically for the truncated fields. Its evolution (not shown) is qualitatively similar to the ones presented here, although weaker and unrelated to the geometric truncation length. The evolution of $E_{uu}(k_x=0)$ is discussed below.  Each line in figure \ref{fig:Lx}(a) represents a distance from the wall, increasing from red to blue. In agreement with figure \ref{fig:spec}, the near-wall structures of $u$ grow quickly to their asymptotic length, while the outer layers grow more slowly and have not reached their final length at the end of the experiment. The elongation speed $\dr L_{uu}/\dr t$ is plotted in figure \ref{fig:Lx}(b), and also has a fast initial transient followed by a slower evolution. Especially interesting are the very high initial celerities near the wall, which are much faster than any velocity fluctuation in the flow. This could perhaps be explained by viscous effects near the wall, but even the logarithmic plateau at early times, $(\dr L_{uu}/\dr t)^\times \sim O(50)$, is too fast for streamwise advection.

Interestingly, the rapid streak elongation only lasts until the initial length increases by a factor of approximately three. This is shown in figure \ref{fig:Lx}(c), which is similar to figure \ref{fig:Lx}(b) but plotted against $L_{uu}$ instead of against time, and is significant because it suggests an alternative mechanism for fast elongation. \cite{jim22_nostr} showed that truncating the streaks does not affect the wall-normal velocity strongly, and figure \ref{fig:spec}(b) shows that $E_{vv}$ is initially active at lengths $\lambda_x/\lambda_f\lesssim 2\mbox{--}3$. We saw in \r{eq:St} that the vertical migration of the velocity fluctuations in figure \ref{fig:cfs}(b) could be explained by a local generation mechanism whose time scale is governed by the local shear, and figure \ref{fig:balu}(a) shows the growth of the long components of the velocities during the initial part of the recovery, integrated over $y$. The streamwise velocity, initially damped by the filter, grows approximately linearly until it almost reaches its equilibrium value, but the transverse velocity components that form the roller change little during that time. Figure \ref{fig:balu}(b) shows the main contributions to the energy equation of the long $u$. Initially, the main contribution is the deformation of the mean shear by the surviving roller (red), as in a regular channel. Later, turbulent transport intensifies, although always weaker than production, and the growth of the streak saturates due to nonlinear interactions with the shorter wavelengths. There are two important contributions to this transport: The positive green line is the self-distortion of the vertical gradient of the streak by the Reynolds stresses, and reinforces the streak. The other component (cyan) is the spanwise advection of the streak by $w$, and damps it \citep{Hamilton95,Waleffe97,Lozano2021}. We will come back to this question when discussing other recovery experiments in \S \ref{sec:short}. These relative contributions are approximately evenly distributed among individual modes, as exemplified in figure \ref{fig:balu}(c) for $k_x=0$. The magnitude of this balance is about 10\% of that in figure \ref{fig:balu}(b), and this is also approximately the fraction of the truncated energy which is associated with $k_x=0$ (not shown). Figure \ref{fig:balu}(d--f) repeats the same information for RS180, illustrating the essential Reynolds-number independence of the described effects.

\begin{figure}
\centerline{%
\raisebox{0mm}{\includegraphics[height=.23\textwidth,clip]%
{ 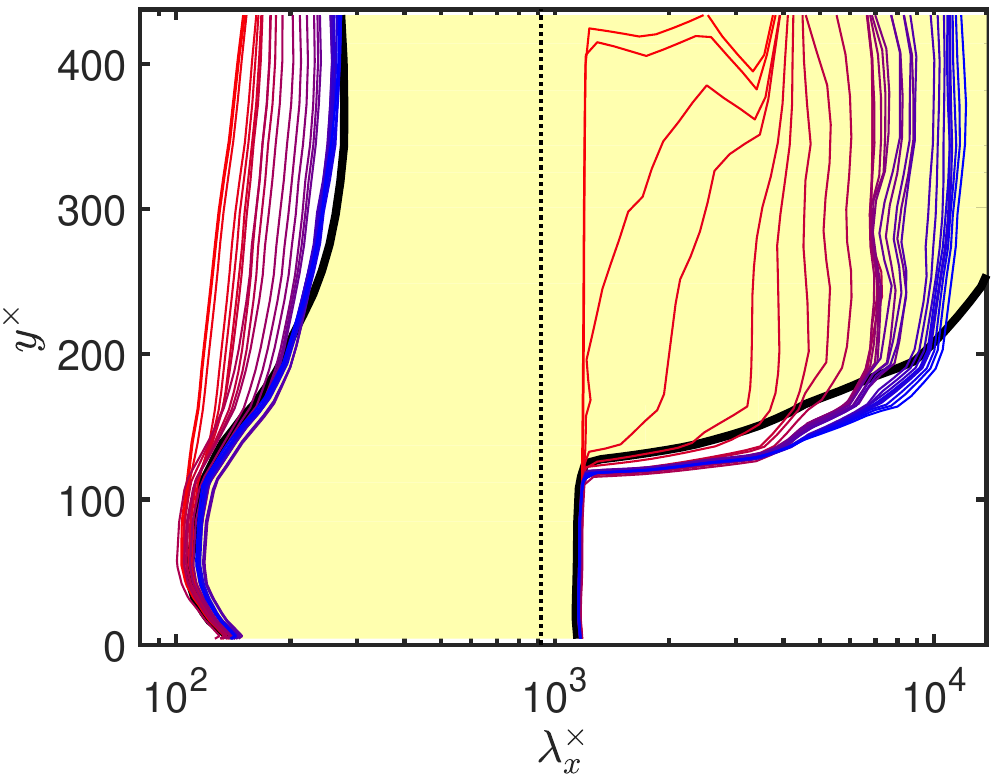}}
\mylab{-.32\textwidth}{.19\textwidth}{(a)}%
\hspace*{1mm}%
\raisebox{0mm}{\includegraphics[height=.23\textwidth,clip]%
{ 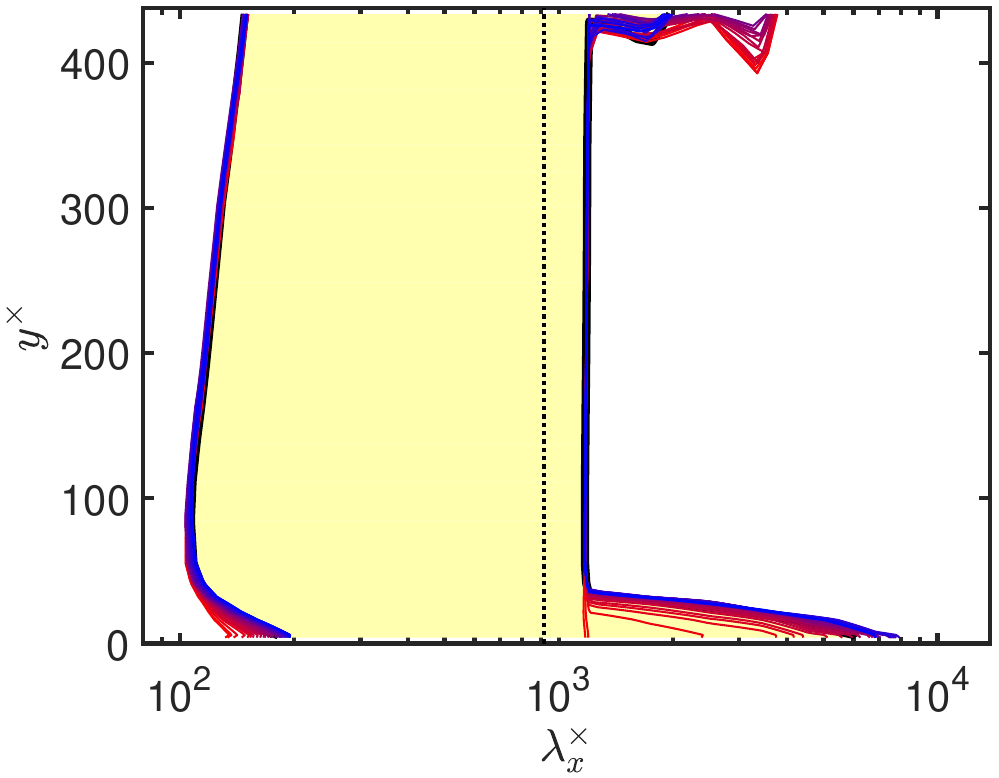}}
\mylab{-.05\textwidth}{.19\textwidth}{(b)}%
\hspace*{1mm}%
\raisebox{0mm}{\includegraphics[height=.235\textwidth,clip]%
{ 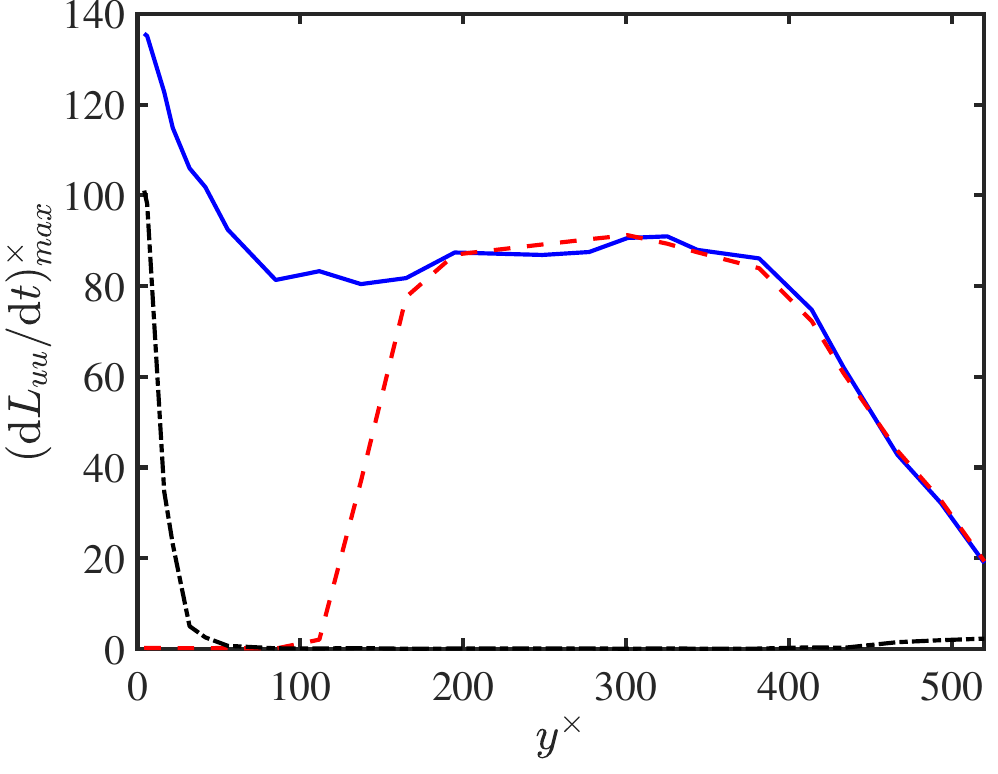}}
\mylab{-.06\textwidth}{.19\textwidth}{(c)}%
}%
\caption{%
(a) Streamwise one-dimensional premultiplied spectrum of the streamwise velocity, $k_x E_{uu}(k_x)/u'^2$, as a function of time and $y$ in RS550:15in. Contours are  $t u_{\tau 0}/h=0(0.013) 0.365$, from red to blue,  and are uniformly scaled with $u_{\tau 0}$ so as contain 90\% of the energy of the spectrum at $t=0$.
The thick black contour with yellow fill is the spectrum of the long-time limit of the partially damped flows, and the vertical dotted line is $\lambda_{f}$ before the release of the damping.
(b) As in (a), for RS550:15out. $t u_{\tau 0}/h=0 (0.015) 0.433$, from red to blue.
(c) Maximum elongation velocity, as a function of wall distance. \solid, RS550:15; \dashed, RS550:15in, as in (a); \chndot, RS550:15out, as in (b). $\alpha=0.85$.
}
\label{fig:Lxinout}
\end{figure}

\subsection{Interaction among layers}\la{sec:intery}

To further test the dependence of the streak growth on the local shear, consider an experiment in which only some distances from the wall are released,
\beq
\homega_y (k_x,y,k_z) \to g(y) \, \homega_y (k_x,y,k_z)
\quad\mbox{for all $k_z$, and}\; k_x<  2\pi/\lambda_{f}.
\la{eq:dampy0}
\eeq
where $g(y)$ is a symmetrised hyperbolic tangent that selects layers above (or below) a given distance from the wall. Figure \ref{fig:Lxinout}(a) shows the spectral evolution when is SS550:15 only released for $y^+\gtrsim 100$ (RS550:15in), while figure \ref{fig:Lxinout}(b) only releases $y^+\lesssim 50$ (RS550:15out).

It is clear that the released layers elongate independently of the unreleased ones. Figure \ref{fig:Lxinout}(c) displays the maximum elongation velocities of the partially released cases, compared to the fully released one from figure \ref{fig:Lx}(c), and shows that the similarity is quantitative. The outer-released case in figure \ref{fig:Lxinout}(a) does not grow for the wall distances that remain damped, but those that have been released grow at the same rate as in the fully released channel. Similarly, the elongation of the inner-released case in figure \ref{fig:Lxinout}(b) is restricted to the released layers near the wall.

Because of this independence among layers, as well as from the relative Reynolds number independence of our results, the rest of the paper will restrict itself to $y^+\lesssim 100$ and $Re_\tau\approx 180$.

\begin{figure}
\centerline{%
\raisebox{0mm}{\includegraphics[height=.22\textwidth,clip]%
{ 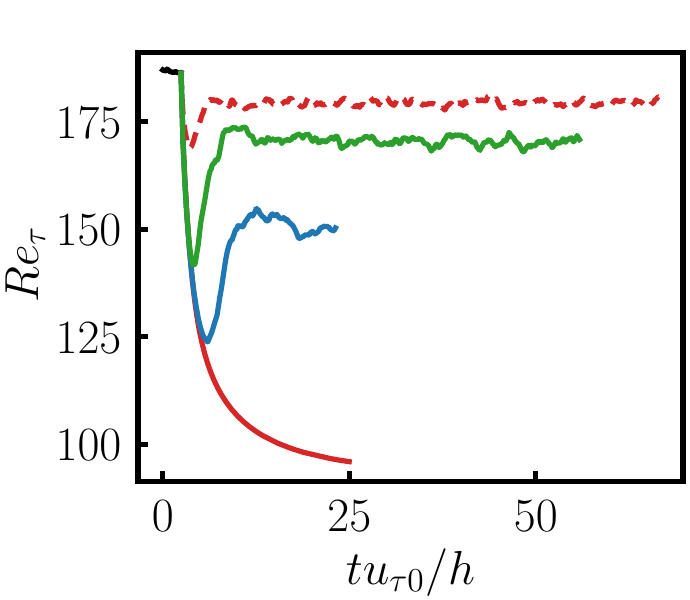}}%
\mylab{-.27\textwidth}{.165\textwidth}{(a)}%
\hspace*{1mm}%
\raisebox{0mm}{\includegraphics[height=.22\textwidth,clip]%
{ 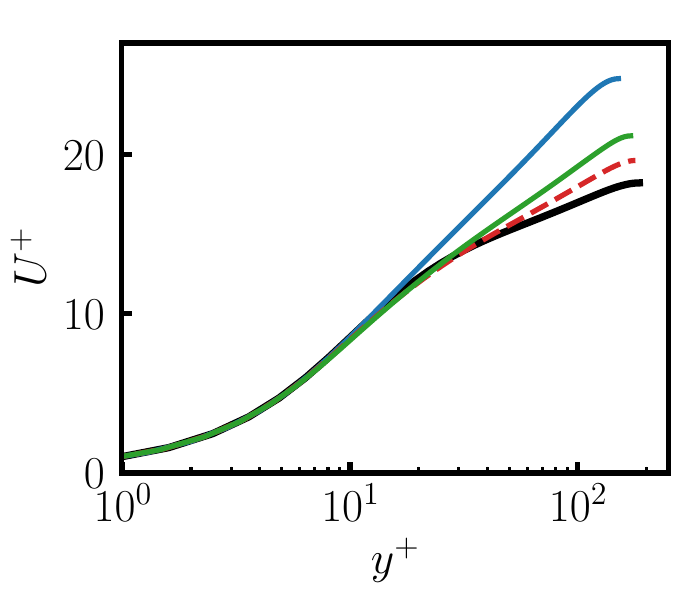}}%
\mylab{-.18\textwidth}{.165\textwidth}{(b)}%
\hspace*{1mm}%
\raisebox{0mm}{\includegraphics[height=.22\textwidth,clip]%
{ 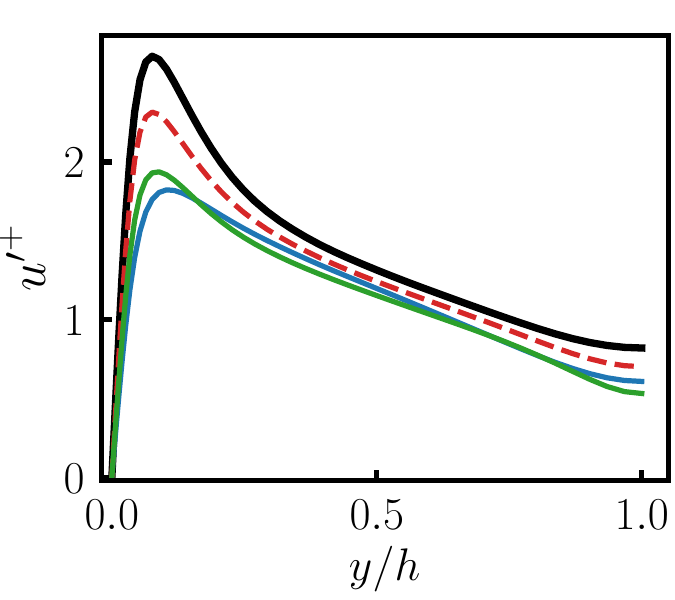}}%
\mylab{-.05\textwidth}{.165\textwidth}{(c)}%
\hspace*{1mm}%
\raisebox{0mm}{\includegraphics[height=.22\textwidth,clip]%
{ 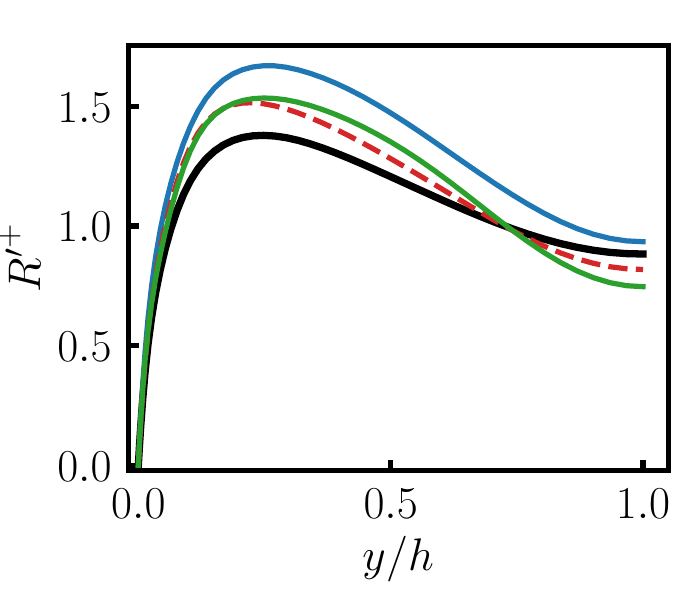}}%
\mylab{-.05\textwidth}{.165\textwidth}{(d)}%
}%
\caption{%
(a) Evolution of $\retau$ for the different damping experiments upon initiation of the damping at $t=0$. 
(b) Mean velocity profile.
(c) Streamwise velocity fluctuations.
(d) Roller intensity. 
Black, reference F180; 
blue, damped streak SS180, $\lambda_f=\Lambda_x/10$; 
dashed red, damped roller SR180:5, $\lambda_f=\Lambda_x/5$;
green, fully damped, SA180, $\lambda_f=\Lambda_x/10$; 
red, damped roller, SR180:10, $\lambda_f=\Lambda_x/10$. 
}
\label{fig:uRprimes}
\end{figure}
 
\section{Damping the rollers}\la{sec:short}

The discussion in the previous section suggests that long rollers, rather than long streaks, are the key ingredient of wall turbulence because, unless inhibited by damping, they deform the mean profile to also build long streaks. However, although the experiments discussed in \S\ref{sec:results} show that streaks are not necessary if rollers are present, they do not prove that the rollers are required. The experiments in this section address this question. The first experiment (SR180) damps the rollers by zeroing the long $\nabla^2 \hv$ without damping the long $\homega_y$ but, since it turns out that this also weakens the streaks, a second experiment (SA180) damps all the long flow structures by zeroing both $\nabla^2\hv$ and $\homega_y$ for $\lambda_x\ge \lambda_f$. Figure \ref{fig:uRprimes}(a) shows that all the truncation experiments initially decay. Some, such as the solid red line of the damped roller SR180:10 $(\lambda_f =\Lambda_x/10)$, eventually laminarise, but in most cases the undamped wavenumbers recover after $tu_{\tau0}/h=O$(5--10), and reach a new stable state. These include a less truncated flow with streaks but no rollers (SR180:5, $\lambda_f =\Lambda_x/5$), and a fully truncated one with $\lambda_f=\Lambda_x/10$ (SA180). Figure \ref{fig:uRprimes}(b-d) displays their velocity profiles after the initial transient. Figure \ref{fig:uRprimes}(b) shows that the mean velocity of all the truncated cases has a stronger wake than a regular channel, implying less efficient Reynolds stresses. They also have weaker streamwise velocities (figure \ref{fig:uRprimes}.c), and a stronger roller (figure \ref{fig:uRprimes}.d). The fluctuations are therefore more isotropic than in a regular flow, no doubt because of the effect of damping the most anisotropic flow scales. Interestingly, when comparing channels with the same cutoff length ($\lambda_f=\Lambda_x/10$), the case closest to the canonical profiles is the fully damped one, SA180. Case SR180:5 is not included in this comparison because its cutoff length had to be increased to avoid laminarisation.

Apparently, keeping one of the two velocity components of the Reynolds stress (streaks or rollers) while damping the other one is less stable than damping both of them. An extra experiment (not shown) was performed applying the same damping as in SA180, but also damping all fluctuations with $\lambda_z>2h$. The flow also recovers, reaching values of $\Rey_\tau$ comparable to those of F180, showing that neither infinitely long or infinitely wide fluctuations are required. This is consistent with the known result that scales wider than $2h$ are not strongly amplified by the shear \citep{ala:jim:06}.

\begin{figure}
\vspace*{0mm}%
\centerline{%
\raisebox{0mm}{\includegraphics[height=.24\textwidth,clip]%
{ 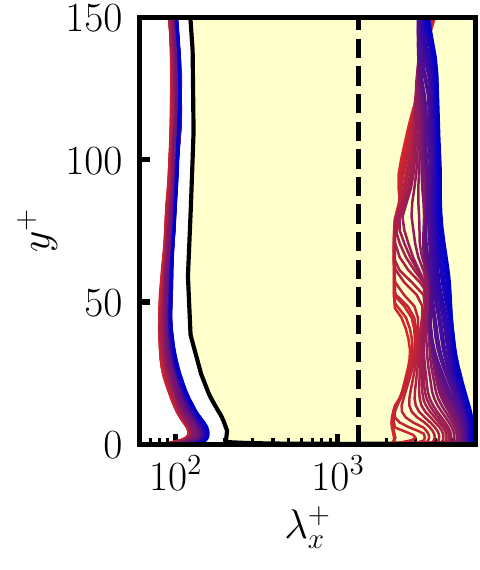}}
\mylab{-.2\textwidth}{.20\textwidth}{(a)}%
\hspace*{3mm}%
\raisebox{0mm}{\includegraphics[height=.24\textwidth,clip]%
{ 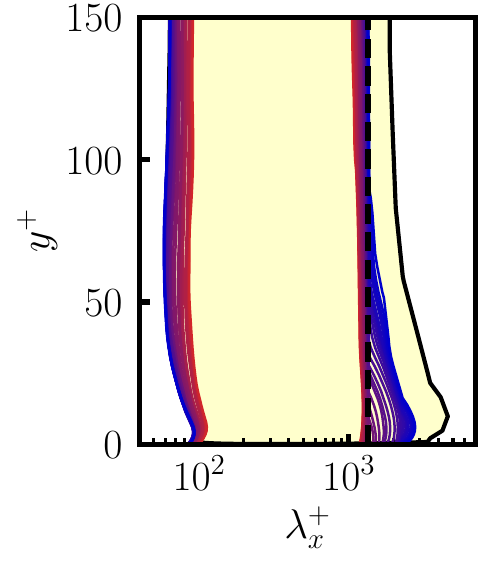}}
\mylab{-.2\textwidth}{.20\textwidth}{(b)}%
\hspace*{1mm}
\raisebox{0mm}{\includegraphics[height=.245\textwidth,clip]%
{ 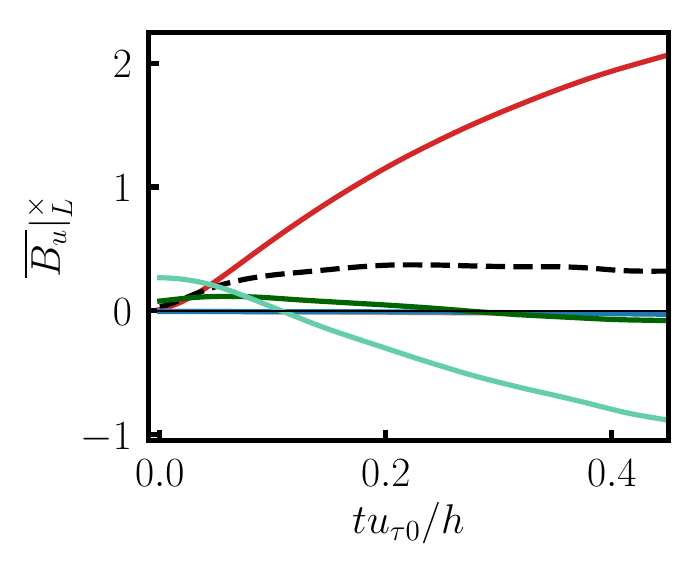}}
\mylab{-.2\textwidth}{.20\textwidth}{(c)}%
\hspace*{2mm}
\raisebox{0mm}{\includegraphics[height=.25\textwidth,clip]%
{ 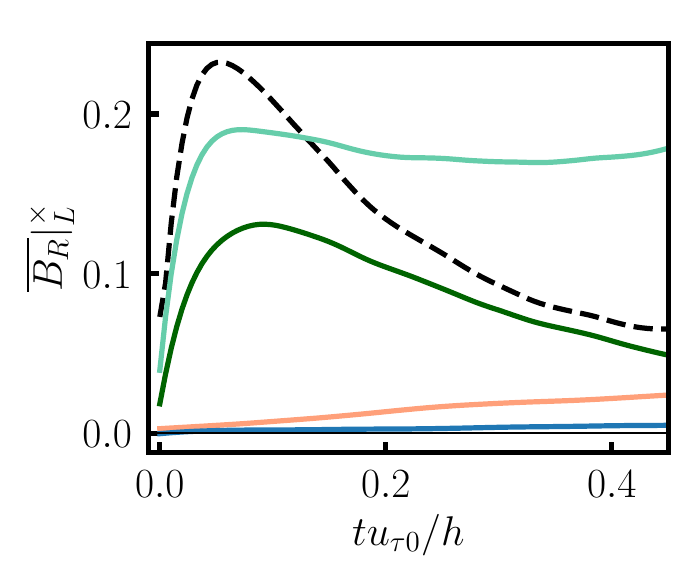}}
\mylab{-.05\textwidth}{.20\textwidth}{(d)}%
}%
\vspace*{1mm}%
\centerline{%
\raisebox{0mm}{\includegraphics[height=.24\textwidth,clip]%
{ 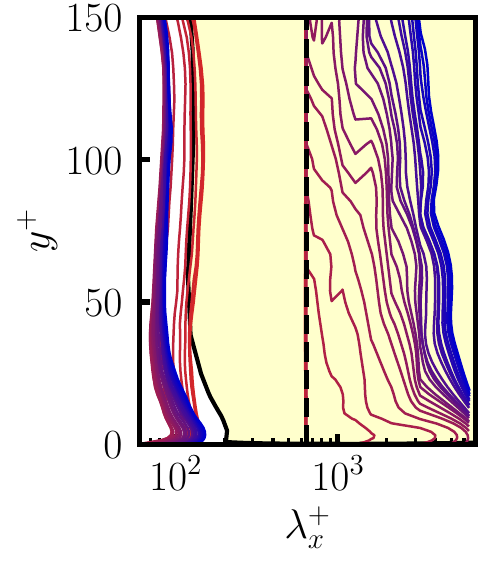}}
\mylab{-.2\textwidth}{.20\textwidth}{(e)}%
\hspace*{3mm}%
\raisebox{0mm}{\includegraphics[height=.24\textwidth,clip]%
{ 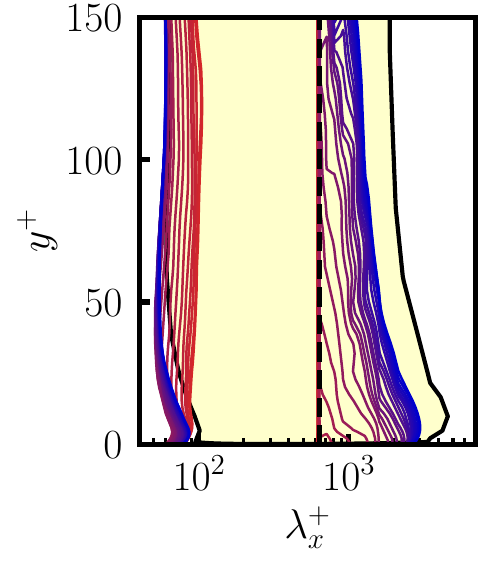}}
\mylab{-.2\textwidth}{.20\textwidth}{(f)}%
\hspace*{.75mm}
\raisebox{0mm}{\includegraphics[height=.25\textwidth,clip]%
{ 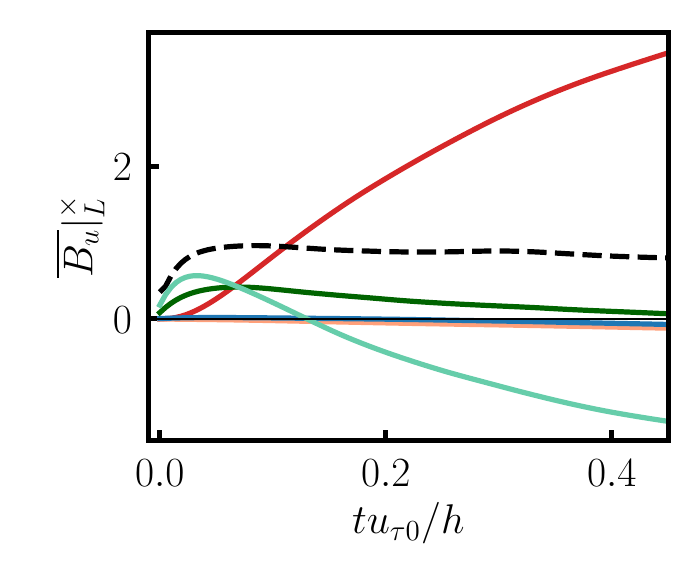}}
\mylab{-.2\textwidth}{.20\textwidth}{(g)}%
\hspace*{1.8mm}
\raisebox{0mm}{\includegraphics[height=.255\textwidth,clip]%
{ 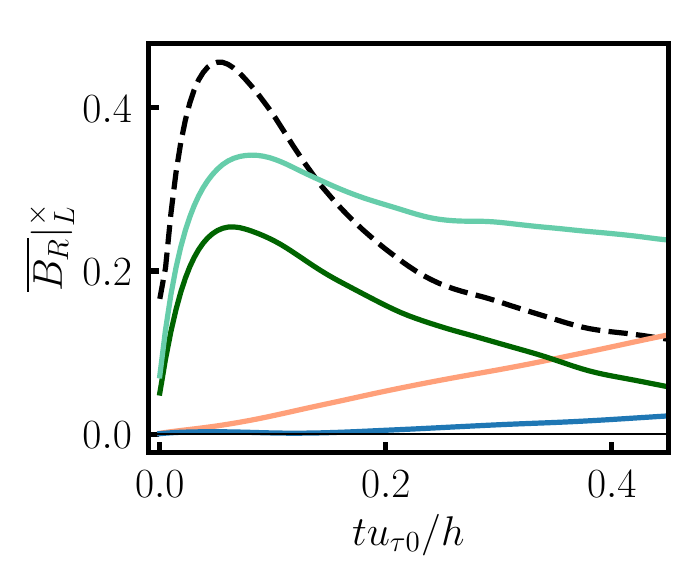}}
\mylab{-.05\textwidth}{.20\textwidth}{(h)}%
}%
\caption{%
(a-d) Damped rollers retaining  long streaks (F180 $\to$ SR180:5  $\to$ RR180).
(a) Evolution of the streak spectrum in the recovery RR180, as in figure \ref{fig:spec}. From red to blue, $tu_{\tau 0}/h=0(0.014) 0.463$.
(b) Roller spectrum.
(c) Energy sources for $u'_L$, as in figure \ref{fig:balu}(b). 
(d) Same for the roller energy.
(e-h) As in (a-d), for fully damped long scales (F180 $\to$ SA180 $\to$ RA180).
$tu_{\tau0}/h=0(.018)0.456$.
}
\label{fig:cfs180}
\end{figure}

Figure \ref{fig:cfs180} presents recovery experiments for these truncated flows. As in \S\ref{sec:results}, two recoveries are run for each flow, and results are averaged over both tests. Figures \ref{fig:cfs180}(a,b) show the evolution of the streak and roller spectrum for the damped-$\nabla^2 v$ experiment, and it is interesting that, although $\omega_y$ has not been damped, the streamwise velocity is also substantially shortened. This case is different from others in this paper because using the standard truncation length, $\lambda_f^\times=\Lambda_x^\times/10 \approx 700$, laminarises the flow. A milder $\lambda_f^\times=1400$ is used in figure \ref{fig:cfs180}, and figure \ref{fig:cfs180}(b) shows that the roller is then barely modified. In addition, gradually decreasing $\lambda_f$ allowed us to reach the standard truncation length, $\Lambda_x/10$, but that flow is very intermittent and is not used in the paper. 

As in the truncated-$\omega_y$ case in \S \ref{sec:results}, the recovery starts from the wall. The roller and the streak grow simultaneously, with the action of the roller on the mean shear initially predominant (figure \ref{fig:cfs180}.c). Note that only the long roller harmonics can initially participate in this essentially linear process, and that only later does the nonlinear transfer become relevant. The roller grows by nonlinear transfer (figure \ref{fig:cfs180}.d). That the rollers and the streaks initially grow simultaneously, with no contribution to the streak generation from the short cross-flow velocities and, as shown by the truncation experiments, with no obvious feedback from the long streaks to the long rollers, again suggests that damping both long components should not kill the flow, and that a fully truncated flow should recover well. We saw in figure \ref{fig:uRprimes} that the first conjecture is true. In fact, that truncation is more robust than only damping the roller. Figures \ref{fig:cfs180}(e-h) document the recovery from the fully damped experiment.

\begin{figure}
\vspace*{5mm}%
\centerline{%
\raisebox{0mm}{\includegraphics[height=.20\textwidth,clip]%
{ 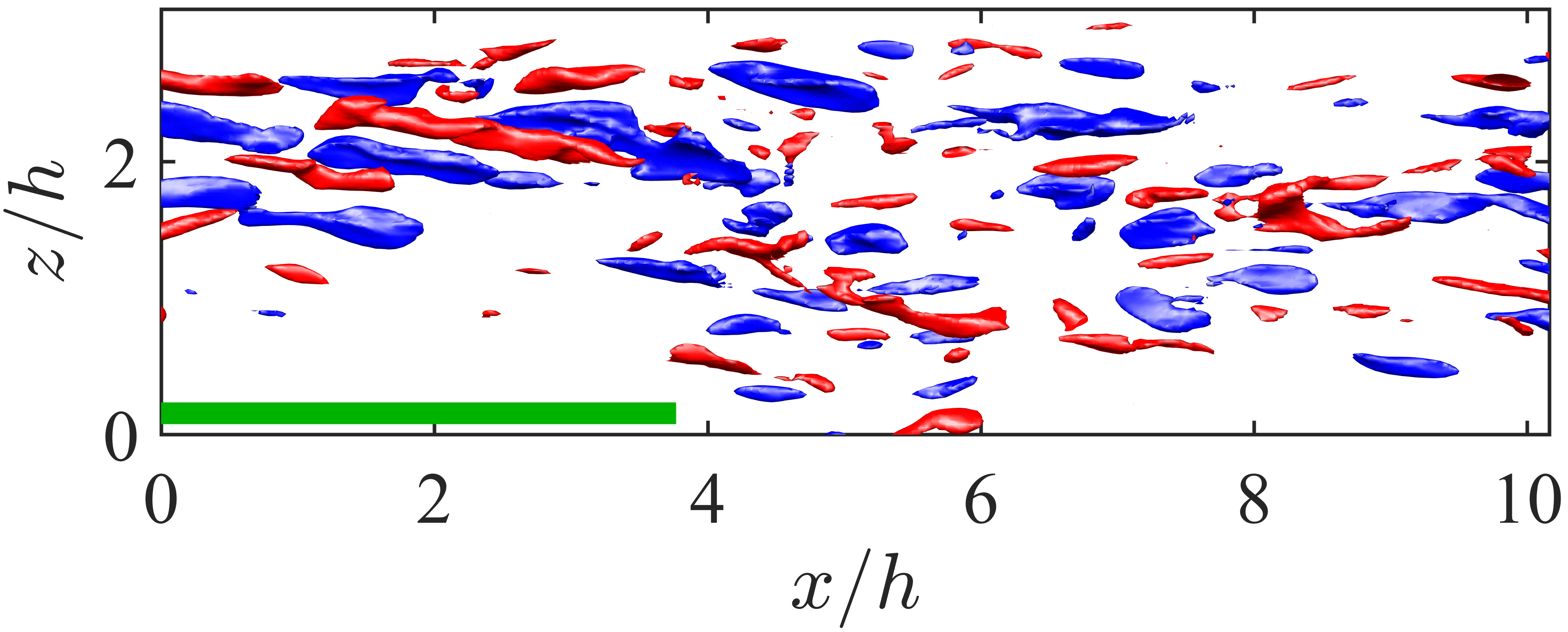}}
\mylab{-.24\textwidth}{.205\textwidth}{(a)}%
\hspace*{2mm}%
\raisebox{0mm}{\includegraphics[height=.20\textwidth,clip]%
{ 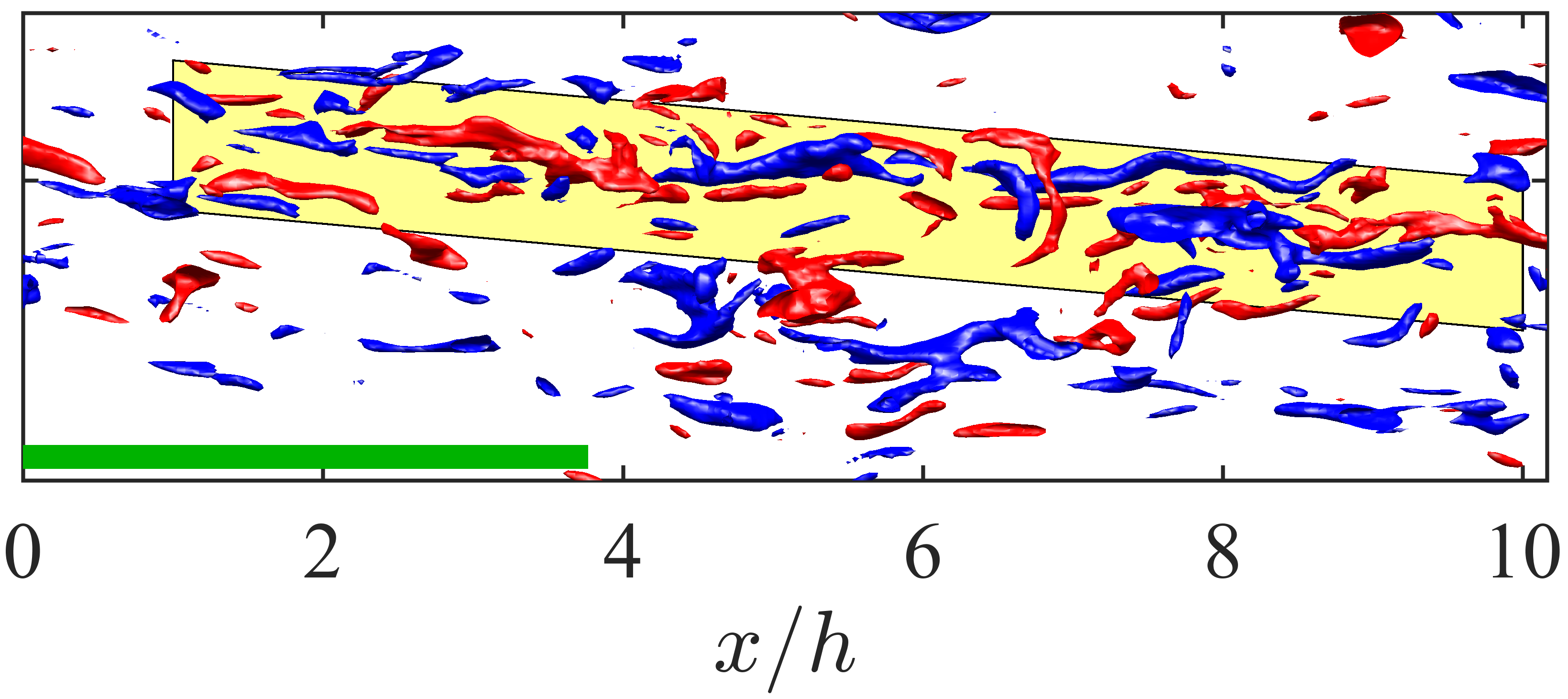}}
\mylab{-.24\textwidth}{.205\textwidth}{(b)}%
}
\caption{%
(a) Streamwise velocity fluctuations, $|u^\times|\ge 3.75$, at $y^\times\le 90$, for the truncated-streak simulation SS180. The green rectangle is the truncation length. Only a small part of the computational box is shown in each direction, and flow is from left to right (b) $|v^\times|\ge 1.75$. Blue is positive; red, negative.
}
\label{fig:planFR}
\end{figure}

Physical fields give a clearer picture of the recovery process. Figure \ref{fig:planFR}(a) displays strong fluctuations of $u$ in the truncated SS180 channel. The long streaks are missing, substituted by shorter segments arranged so that their average over lengths longer than $\lambda_f$ vanishes. The length of individual segments, $\ell_x^\times \approx 400$, approximately agrees with the one of strong $u$- or $v$-structures in regular channels \citep{jim18}. We saw in figure \ref{fig:spec}(b) that the spectrum of $v$ retains energy beyond the truncation wavelength of $u$, but the $v$-structures in figure \ref{fig:planFR}(b) are not noticeably longer than those of $u$. However, they are arranged into groups whose $x$-average does not cancel (see, for example, the highlighted yellow trail in the upper half of the figure, in which structures are predominantly positive). \cite{Hamilton95} studied the formation of $k_x=0$ rollers during the natural oscillations of their minimal Couette flow, and concluded that they are sustained by nonlinear mode interactions, generally involving $\hu$, but incorporating only contributions from shorter wavenumbers. Adapting their analysis to our case, the picture is of shorter structures organising themselves into groups that populate the long end of the spectrum. The above truncation experiments suggest that these groups do not feed back into the shorter wavelengths, which thus constitute an essentially autonomous turbulence `engine'.

\section{Conclusions}\la{sec:conc}

This paper continues the damping experiments in \cite{jim22_nostr}, which showed that the long streaks of the streamwise velocity in wall-bounded flows are essentially passive products of shorter structures with $\lambda_x^+\lesssim 600$. We have first performed recovery experiments in which the damping is removed, and shown that the streaks are predominantly recreated by the effect of residual long `rollers' of the transverse velocities. This motivates us to damp the long rollers, first while retaining long streaks, and then by damping all long fluctuations. Turbulence survives in both cases, implying that, although many scales are present in unconstrained flows, and several sustenance mechanisms may coexist, wall turbulence can function with only short velocity fluctuations. Recovery experiments of these fully truncated flows show that long rollers are maintained by nonlinear turbulent transport from shorter units, while the streaks reform by the linear effect of the rollers on the mean shear and their growth is saturated by the damping effect of nonlinear transport.

Note that the new truncated flows are different from the minimal units in \cite{jim:moi:91}, which retain infinitely long structures. It was hypothesised by \cite{lozano14} that these features mimic the effect of long but finite structures in natural flows, but even the infinite component is absent in the new short units.

\acknowledgements
This work was supported by the European Research Council under the Caust grant ERC-AdG-101018287. JJ would like to acknowledge discussions with Julio Soria regarding the experiments  in figure \ref{fig:Lxinout}.
\\
\\
Declaration of Interests. The authors report no conflict of interest.
%

\end{document}